\documentclass[prl,aps,amsmath,twocolumn,floatfix,showpacs,superscriptaddress]{revtex4}

\usepackage{bm}
\usepackage{graphicx}
\usepackage{multirow}
\usepackage{natbib}

\begin{document}

\title{Geometric phase gate on an optical transition for ion trap quantum computation}

\author{K.~Kim}
\affiliation{Institut f{\"u}r Experimentalphysik, Universit{\"a}t
Innsbruck, Technikerstrasse 25, A-6020 Innsbruck, Austria}

\author{C.~F.~Roos}
\affiliation{Institut f\"ur Quantenoptik und Quanteninformation der
\"Osterreichischen Akademie der Wissenschaften, Technikerstr. 21a,
A-6020 Innsbruck}

\author{L.~Aolita}
\affiliation{Instituto de F\'\i sica, Universidade Federal do Rio de
Janeiro, Caixa Postal 68528, 21941--972 Rio de Janeiro, RJ, Brazil}

\author{H.~H\"affner}
\author{V.~Nebendahl}
\affiliation{Institut f\"ur Quantenoptik und Quanteninformation der
\"Osterreichischen Akademie der Wissenschaften, Technikerstr. 21a,
A-6020 Innsbruck}

\author{R. Blatt}
\affiliation{Institut f{\"u}r Experimentalphysik, Universit{\"a}t
Innsbruck, Technikerstrasse 25, A-6020 Innsbruck, Austria}

\affiliation{Institut f\"ur Quantenoptik und Quanteninformation der
\"Osterreichischen Akademie der Wissenschaften, Technikerstr. 21a,
A-6020 Innsbruck}

\date{\today}

\begin{abstract}
We propose a geometric phase gate of two ion qubits that are encoded
in two levels linked by an optical dipole-forbidden transition.
 Compared to hyperfine geometric phase gates
mediated by electric dipole transitions, the gate has many
interesting properties, such as very low spontaneous emission rates,
applicability to magnetic field insensitive states, and use of a
co-propagating laser beam geometry. We estimate that current
technology allows for infidelities of around 10$^{-4}$.
\end{abstract}


\maketitle One of the important and most difficult experimental
efforts of quantum computation is to realize almost perfect
two-qubit gate operations. Currently, it is believed that
 gate error probabilities of about 10$^{-4}$ would be sufficiently low to allow for
so-called efficient fault-tolerant quantum computing.
\cite{Knill,Reichardt}. Strings of trapped ions are among the most
promising candidates for the realization of a quantum computer. So
far, the lowest published gate infidelity is still around 3\%
\cite{Leibfried} and was experimentally achieved with a geometric
phase gate in an ion trap experiment. The main limitations of this
 gate come from spontaneous emission and magnetic field
fluctuations \cite{Leibfried, Lee}.

Ion trap quantum computation can be implemented with two
alternative qubit encodings: hyperfine ground state qubits and
qubit states connected by optical transitions. For hyperfine
qubits, the gate operations are performed by Raman coupling
mediated by dipole transitions. Ref. \cite{Leibfried} used an
encoding based on such a hyperfine transition. In this setting,
however, it is demanding to reduce spontaneous scattering below
the required fault tolerant level \cite{Ozeri, OzeriA}, because a
tremendous amount of laser power is required. Recently, the use of
Raman processes on quadrupole transitions was proposed for
hyperfine qubits \cite{Aolita2}. However, also this strategy
requires high laser powers to achieve short gate times.

Here, we propose a $\sigma^z$-type geometric phase gate on an
optical transition to overcome some of the limitations present in
the realization of \cite{Leibfried}. For instance the use of an
optical quadrupole transition allows to reduce the likelihood of a
spontaneous emission event sufficiently. Also it is shown that
magnetic field insensitive states can be used for the $\sigma^z$
geometric gate on an optical transition. More interestingly, the
gate can be executed with a co-propagating laser beam configuration,
which reduces the errors from phase fluctuations between two laser
beams \cite{Lee}. With hyperfine qubits, on the contrary, only a
counter-propagating scheme can be utilized for the gate. Finally,
the proposed gate can be directly used as a logical two-qubit gate
in dephasing-free subspaces because of the natural suitability of
phase gates for such purposes \cite{Aolita1}.

The gate here proposed applies to any ion-qubit states connected by
weak transitions such as, quadrupole transitions of Ca${^+}$,
Sr${^+}$, Ba${^+}$. We first show that it is possible to realize a
state-dependent displacement on the optical transitions with
bichromatic laser radiation, and that for the most interesting detunings of the laser fields the
 coupling is maximized in a
co-propagating geometry. In turn, the applicability of the gate to
magnetic field insensitive states is explained. In a second step,
we extend the scheme to two ions and study carefully the intrinsic
complications leading to infidelities of the gate. We show that
these can be compensated by spin echo techniques, thus reducing
the infidelities to a level of about 10$^{-4}$. We also discuss
the connection between gate speed and the probability of
spontaneous emissions during one gate operation, and show that the
error due to it can also be reduced to a level of 10$^{-4}$.
Finally, we briefly examine other more technically relevant errors
arising from fluctuations of experimental parameters.

Consider a single ion in a one-dimensional harmonic trap
interacting with two laser beams detuned from the ion's quadrupole
transition connecting the ground state $S$ to a metastable $D$
state. If we treat the ion as a two-level system, the Hamiltonian
in the interaction picture and after performing the rotating wave
approximation with respect to optical frequencies is given by
\begin{eqnarray}
\hat{H}_{I}=\sum_{j=1,2} \hat{\sigma}^{+} ( \frac{\hbar
\Omega_{j}}{2} e^{-i (\Delta_j t +\phi_j)} e^{i \eta_j (\hat{a}
e^{-i \nu t} + \hat{a}^{\dag} e^{i \nu t})}) + h.c.,
\label{startingHam}
\end{eqnarray}
where $\hat{\sigma}^{+} = |D\rangle \langle S|$, $\hat{a}$ and
$\hat{a}^{\dag}$ are the ladder operators of the oscillator, $\nu$
is the trap frequency, and $\Omega_{j}$, $\eta_{j}$ are the Rabi
frequency and Lamb-Dicke parameter of the laser with detuning
$\Delta_{j}$ and optical phase $\phi_{j}$, respectively.

As will be detailed below, a state-dependent displacement
operation is achieved by setting $\Delta_{1}=\Delta$,
$\Delta_{2}=\Delta-\nu+\delta$ ( $\delta \ll \nu$). In the
Lamb-Dicke regime, at low laser intensity $\Omega_{j}\ll
\Delta_{j}$, and ignoring terms faster than $\delta$, a
second-order perturbation yields the following effective
interaction Hamiltonian
\begin{eqnarray}
\hat{H}_{\rm eff,1}&=& (\frac{\hbar \Omega_{\rm eff}}{2} \hat{a}
e^{-i \delta t }+ \frac{\hbar \Omega_{\rm eff}^{*}}{2}
\hat{a}^{\dag} e^{i \delta t }) \hat{\sigma}^{z} + {\rm LS}.
\label{effectiveHamforone}
\end{eqnarray}
Here, $\Omega_{\rm eff} = ( \frac{\eta_{1}
(\Delta-\nu+\delta/2)}{(\Delta-\nu)(\Delta-\nu+\delta)}-
\frac{\eta_{2} (\Delta+\delta/2)}{\Delta (\Delta+\delta)})
\frac{\Omega_{1} \Omega_{2}}{2} e^{-i \phi_L}$, and ${\rm LS}$
denotes the light shifts coming from the transitions of carrier
and the first motional sidebands and is given by $\sum_{j=1,2}
\hbar \Omega_{j}^2 \hat{\sigma}^{z} [- \frac{1}{4 \Delta_{j}} -
(\frac{\eta_{j}^2}{4 (\Delta_{j}+\nu)}- \frac{\eta_{j}^2}{4
(\Delta_{j}-\nu)})(\hat{n}+\frac{1}{2})]$, where $\hat{\sigma}^{z}
= |D\rangle \langle D|-|S\rangle \langle S|$, $\hat{n}=
\hat{a}^{\dag} \hat{a}$, $\phi_L=\phi_1-\phi_2-\pi/2$. Neglecting
light shifts for the moment, the effective
Hamiltonian~(\ref{effectiveHamforone}) describes the desired state
dependent displacement operation \cite{Lee}. The time evolution
operator is found to be
\begin{eqnarray}
\hat{U}(t)&=&\exp\{-\frac{i}{\hbar} (\int_{0}^{t} dt' \hat{H}_{\rm
eff}(t') \label{calculationoftimeevolution} \\ &-&
\frac{i}{2\hbar}\int_{0}^{t} dt'\int_{0}^{t'} dt'' [\hat{H}_{\rm
eff}(t'),\hat{H}_{\rm eff}(t'')]+\cdots)\} \nonumber ,
\end{eqnarray}
from the Magnus expansion related to the Baker-Campbell-Hausdorff
formula $e^{A}e^{B} = \exp(A + B + \frac{1}{2} [A, B] + \cdots)$.
With the Eq. (\ref{calculationoftimeevolution}), the time
evolution operator of the Hamiltonian ~(\ref{effectiveHamforone})
can be obtained by $\hat{U_{1}}(t) = e^{(\alpha(t)
\hat{a}^{\dagger}-\alpha(t)^{*} \hat{a}) \hat{\sigma}_{z} } e^{i
\Phi(t) \hat{\sigma}_{z}^2} $. Here, $\alpha(t) =
\frac{\Omega_{\rm eff}}{2 \delta}(1 - e^{i \delta t})$ and
$\Phi(t)= |\frac{\Omega_{\rm eff}}{2 \delta}|^{2} (\delta t - \sin
(\delta t))$. The ion moves periodically along a circular path in
phase space of radius $|\hbar \Omega_{\rm eff}/ 2 \delta|$  with
periodicity $2\pi/\delta$, and the direction of motion is
determined by the qubit state and  $\phi_L$. At $t=2 \pi/\delta$,
it returns to the original motional state and acquires the
geometrical phase $\Phi_{g}= 2 \pi |\Omega_{\rm eff}/{2
\delta}|^2$.  The latter phase depends only on the area enclosed
by the trajectory, so both qubit states gain the same geometrical
phase independently of $\phi_L$.
\begin{figure}[hb]
\includegraphics[width=0.9\linewidth]{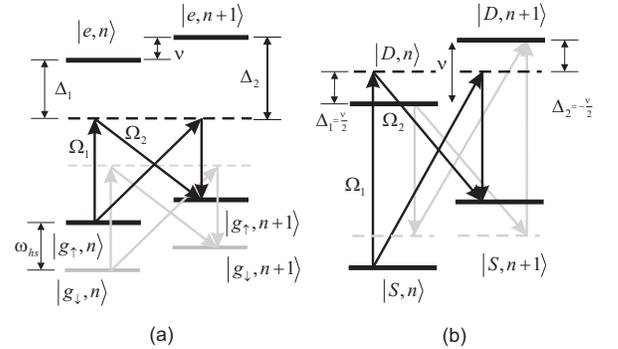}
\caption{\label{fig:copropagating} Schematic representation of
bichromatic laser frequencies to excite the motional states from
$|n\rangle$ to $|n+1\rangle$ (a) for hyperfine qubit states
connected via dipole transitions and (b) for states of the
optical-transition qubit. To keep the figure simple, we omitted
similar couplings connecting $|n-1\rangle$ to $|n-2\rangle$,
$|n+1\rangle$ to $|n+2\rangle$ and so on. On resonance, this
ladder of interactions produces a displacement operation in the
motional state space. In (a), $|g_{\downarrow,\uparrow}\rangle$
and $|e\rangle$ stand for two hyperfine ground states and an
electric-dipole excited state, respectively.}
\end{figure}

We show now that, for optical-transition qubits, a co-propagating
geometry maximizes the strength of the Raman coupling $\Omega_{\rm
eff}$. For the case of hyperfine ground state qubits connected by
dipole transitions, the detunings $\Delta_{1}$, $\Delta_{2}$ must
be much larger than $\nu$ in Fig. \ref{fig:copropagating}(a), so
$\Delta_{1} \simeq \Delta_{2} = \Delta$, which implies that
$|\Omega_{\rm eff}| = |\frac{\eta_{1}-\eta_{2}}{\Delta}
|\frac{\Omega_{1} \Omega_{2}}{2}$. It is then essential to use a
non-copropagating laser beam configuration
($\eta_{1}\neq\eta_{2}$) in order to achieve a non-vanishing
coupling. With quadrupole transitions, the detunings can be of the
order of $\nu$ without considerable spontaneous emission. At the
detunings $\Delta_{1}=-\Delta_{2}\approx \frac{\nu}{2}$, the
coupling strength $|\Omega_{\rm eff}|$ =
$|-\frac{\eta_{1}}{\Delta_{1}}+\frac{\eta_{2}}{\Delta_{2}}
|\frac{\Omega_{1}\Omega_{2}}{2}$ is maximized at
$\eta_{1}=\eta_{2}$. The Raman coupling with those detunings are
depicted in Fig. \ref{fig:copropagating}(b). Most interestingly,
the co-propagating geometry reduces optical phase fluctuations
from path instabilities. Furthermore the co-propagationg geometry
also ensures that the displacement operation can be executed
regardless of the ions' spacing. This is in contrast to a
counter-propagating geometry where it is necessary to carefully
control such spacings so as to have the proper laser phase on each
ion \cite{Leibfried}.

Moreover, the symmetry of the detunings guarantees that the light
shift in Eq.~(\ref{effectiveHamforone}) disappears provided that
both lasers' intensities coincide $\Omega_1=\Omega_2$. Thus, it is
not necessary to consider polarization states to equalize AC stark
shifts of internal states from the two laser beams. Finally, the
state-dependent coupling is achieved without any restriction on the
magnetic-field properties of the states. The scheme here proposed
is, therefore, applicable to magnetic field insensitive transitions,
e.g. the quadrupole transitions of $^{43}$Ca$^{+}$ ion
\cite{Benhelm}.

Now we extend the above consideration to two ions and study the two
qubit gate operation with the detunings
$\Delta_{1}=\frac{\nu}{2}-\frac{\delta}{2}$,
$\Delta_{2}=-\frac{\nu}{2}+\frac{\delta}{2}$ and the same Rabi
frequency $\Omega_{1}=\Omega_{2}=\Omega$. We focus on the center of
mass mode (CM). With two ions, the effective Hamiltonian is given by
\begin{eqnarray}
\hat{H}_{\rm eff,2}&=& \frac{\hbar |\Omega_{\rm eff}|}{2}( \hat{a}
e^{-i (\delta t + \phi_{L}) }+ \hat{a}^{\dag} e^{i (\delta t +
\phi_{L})}) \hat{S}^{z} \nonumber \\ &+& \frac{4}{3} \frac{\hbar
\eta |\Omega_{\rm eff}|}{2} (\hat{\sigma}^{-\it{\Delta} \varphi/
\rm{ 2}}_{1} \otimes \hat{\sigma}^{\it{\Delta} \varphi/ \rm{
2}}_{2}) + \hat{O}(\eta^{3}), \label{effectiveHamfortwo}
\end{eqnarray}
where $|\Omega_{\rm eff}|= \frac{2 \eta \Omega^{2}}{\nu}$,
$\hat{S}^{z}=\hat{\sigma}^{z}_{1} + \hat{\sigma}^{z}_{2}$,
$\hat{\sigma}^{\varphi}_{j} = e^{i \varphi} \hat{\sigma}^{+}_{j}+
e^{-i \phi} \hat{\sigma}^{-}_{j}$, $\it{\Delta} \varphi = k_{L}
\cos{ \theta}~(z_{\rm 1}-z_{\rm 2})$ is the laser phase difference
between both ions, where $k_{L}$ the wave vector, $\theta$ the angle
between $k_{L}$ and trap axis, and $z_{j}$ the equilibrium position
of the $j$th ion. The
first term of Eq. (\ref{effectiveHamfortwo}) produces the evolution of a two-ion geometric phase gate: 
electronic states $|DD\rangle, |SS\rangle$ get both a $\Phi_{g} =
2 \pi |\Omega_{\rm eff}/{\delta}|^2$ phase after $t=2\pi/\delta$,
whereas  states $|DS\rangle, |SD\rangle$ are not affected. At
$\delta= \pm 2 \Omega_{\rm eff}$, the obtained geometric phase
$\Phi_{g}$ must be $\pi/2$ for the gate to be maximally
entangling.

\begin{figure}
\includegraphics[width=0.9\linewidth]{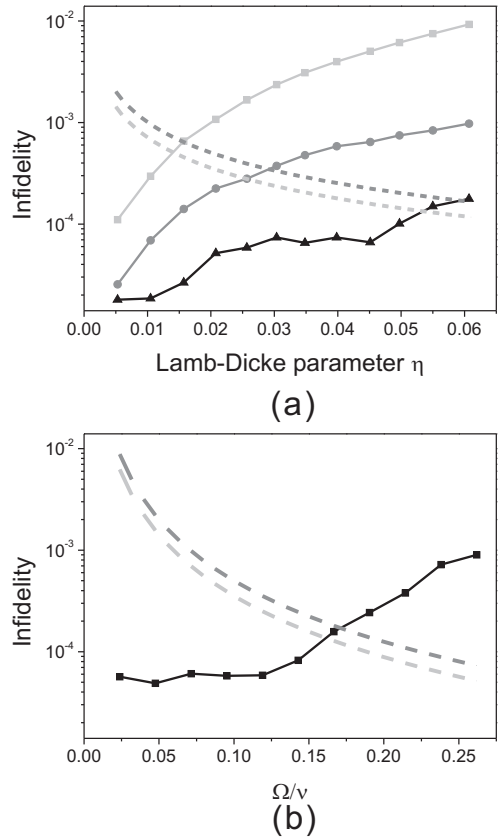}
\caption{\label{fig:etaomegadependency} (a) Gate error vs.
Lamb-Dicke factor $\eta$. The solid lines are the infidelities
obtained from a numerical integration of the full Hamiltonian. The
light gray line is obtained without spin-echo sequence, the gray
one with spin-echo sequence and the black one with spin-echo
sequence and offset detuning $\delta_{\rm off}= (2 \pi) 20$kHz.
Here $\Omega=\nu/6$, $\nu=(2 \pi)1.26$MHz are used. The dominant
error is caused by the MS coupling term. The dashed lines are
infidelities caused by spontaneous decay of the metastable D state
(in the case of Ca$^{+}$). The light gray curve is without
spin-echo and gray curve includes a spin-echo sequence. The
infidelities from spontaneous emissions are determined by the
ratio between gate operation time and life time. (b) Gates error
vs Rabi frequency. The black solid line is the infidelity as a
function of Rabi frequency $\Omega$ with 10 $\mu$s pulse shaping,
which is $10 \sim 100$ times smaller than the gate time. The
integration is executed with spin-echo sequence and $\delta_{\rm
off}= (2 \pi) 20$kHz at $\eta=0.056$, $\nu=(2 \pi)1.26$MHz. Here
the results are obtained by carefully controlling the phase difference
between two lasers at the second pulse of the geometric phase gate
to reduce an order of 0.1\% errors. As in (a), dashed lines indicate
errors caused by spontaneous emission}
\end{figure}

The second term in Eq. (\ref{effectiveHamfortwo}) is a
M{\o}lmer-S{\o}rensen (MS) coupling \cite{Molmer-Sorensen}, which
does not occur for the geometric phase gate with hyperfine qubits.
During the gate operation, the coupling transfers the population
between $|DD\rangle$ and $|SS\rangle$, and between $|DS\rangle$
and $|SD\rangle$. Since populations do not change in an ideal
$\sigma^z$ geometric phase gate, the MS coupling increases the
gate infidelity by producing population errors. We study the
effect of MS coupling by a numerical integration of the exact
Hamiltonian (\ref{startingHam}) of two ions. We sandwich the gate
with two local pulses to turn it into a CNOT gate and assess the
performance of the gate through the state fidelity, the overlap
between the final and the ideal states, using $|DD\rangle$ as the
input state. All other input states display the same fidelities.
As shown in the light-gray solid curve of Fig.
\ref{fig:etaomegadependency}(a), the infidelities are increased to
1\% around $\eta$=0.06 at the given simulation parameters mainly
due to the MS coupling. Including MS coupling and keeping terms up
to $\eta^2$ inside the exponent of Eq.
(\ref{calculationoftimeevolution}), the time evolution operator
for $t=2\pi/\delta$ and $\delta=2 \Omega_{\rm eff}$ can be
approximated by
\begin{eqnarray}
\hat{U}_{2}^{\rm gate}\approx e^{i \frac{\pi}{2}
(\frac{\hat{S}^{z}}{2})^{2}} e^{ -i \frac{2 \pi \eta }{3}
(\hat{\sigma}^{-\it{\Delta} \varphi/\rm{2}}_{1} \otimes
\hat{\sigma}^{\it{\Delta} \varphi/\rm{2}}_{2})}.
\label{Timeevolution}
\end{eqnarray}
Here, we can see that the MS coupling changes the population of
the final state at the end of the gate by $(\frac{2\pi
\eta}{3})^2$, in agreement with the light gray solid curve of Fig.
\ref{fig:etaomegadependency}(a).

Using a spin-echo sequence and an offset detuning $\delta_{\rm
off}$, however, the MS coupling in the gate can be reduced. The
spin-echo sequence exchanges the populations of internal
electronic states of both ions $[e^{i \frac{\pi}{2}
(\hat{\sigma}^{y}_{1} + \hat{\sigma}^{y}_{2})}]$ at the middle and
at the end of the full gate operation. The $\hat{\sigma}^z$ gate
is divided in two parts and at each gate pulse, the same
electronic states get a $\pi/4$ phase, after performing a closed
circle phase space with the radius reduced by $\sqrt{2}$.
Therefore, one needs to increase $\delta = 2 \sqrt{2} \Omega_{\rm
eff}$ for the spin-echo sequence. Around the end of one gate
pulse, the time evolution operator can be approximated by
$\hat{U}_{2}^{\rm echo}= e^{i \frac{\pi}{4}
(\frac{\hat{S}^{z}}{2})^{2}} e^{-i \frac{2 \pi \eta }{3 \sqrt{2}}
(\hat{\sigma}^{-\it{\Delta} \varphi/\rm{2}}_{1} \otimes
\hat{\sigma}^{\it{\Delta} \varphi/\rm{2}}_{2}) }$. We have to add
a $\pi/2$ phase to one ion $[e^{i
\frac{\pi}{4}\hat{\sigma}^{z}_{\rm{1}} }]$, in order to cancel out
the MS operation. This can be shown by the following calculation:
\begin{eqnarray}
e^{i \frac{\pi}{2} (\frac{\hat{S}^{z}}{2})^{2}}=e^{i \frac{\pi}{2}
(\hat{\sigma}^{y}_{1} + \hat{\sigma}^{y}_{2})} \hat{U}_{2}^{\rm
echo} e^{i \frac{\pi}{2} (\hat{\sigma}^{y}_{1} +
\hat{\sigma}^{y}_{2})} e^{i \frac{\pi}{4}\hat{\sigma}^{z}_{\rm{1}}
} \hat{U}_{2}^{\rm echo} . \label{spinechoeffect}
\end{eqnarray}
For introducing a $\pi/2$ phase shift, either a single off-resonant
laser can be used or the distance between the ions can be changed.
In Fig. \ref{fig:etaomegadependency}(a), the gray solid curve shows
that the spin-echo sequence reduces, indeed, the gate infidelity due
to MS coupling.

Eq. (\ref{Timeevolution}) and (\ref{spinechoeffect}), however, are
an approximation up to the order of $\eta^2$ inside the exponent
of Eq. (\ref{calculationoftimeevolution}). Since $\hat{S}^z$ does
not commute with $\hat{\sigma}^{-\it{\Delta} \varphi/\rm{2}}_{1}
\otimes \hat{\sigma}^{\it{\Delta} \varphi/\rm{2}}_{2}$, there is a
term $[\hat{S}^z,\hat{\sigma}^{-\it{\Delta} \varphi/\rm{2}}_{1}
\otimes \hat{\sigma}^{\it{\Delta}
\varphi/\rm{2}}_{2}]\propto\eta^3$ in the time evolution of Eq.
(\ref{calculationoftimeevolution}). Those terms cannot be
compensated by spin-echo pulses. In order to reduce the effect, we
can add a detuning $\delta_{\rm off}$ to both laser frequencies.
The offset detuning $\delta_{\rm off}$ makes the coupling between
$|DD\rangle$ and $|SS\rangle$ off-resonant by $2 \delta_{\rm
off}$, since the same electronic states are connected through red
detuned and blue detuned lasers. Only the population transfer
between $|DS\rangle$ and $|SD\rangle$ is therefore resonant and
dominant in MS coupling. Taking only resonant terms, the MS
coupling in Eq. (\ref{effectiveHamfortwo}) leads to $\frac{4}{3}
\hbar \eta |\Omega_{\rm eff}| (\hat{\sigma}^{+}_{1} \otimes
\hat{\sigma}^{-}_{2} e^{i \it{\Delta} \varphi})+h.c.$ due to the
offset detuning $\delta_{\rm off}$. Since $\hat{\sigma}^{+}_{1}
\otimes \hat{\sigma}^{-}_{2} e^{i \it{\Delta} \varphi}$ commutes
with ${\hat{S}^z} $, Eqs. (\ref{Timeevolution}) and
(\ref{spinechoeffect}) are valid up to order $\eta^3$ inside the
exponent of Eq. (\ref{calculationoftimeevolution}). The induced
phase error from adding $\delta_{\rm off}$ is compensated by the
spin-echo sequence. The black solid curve in Fig.
\ref{fig:etaomegadependency}(a) shows that including the offset
detuning $\delta_{\rm off}$ lowers the error to around 10$^{-4}$
order even at large $\eta$.

In order to reduce the spontaneous emission during the gate, fast
gate operation is important especially for optical-transition
qubits. For the meta-stable states, the spontaneous emission
probability $P_{\rm sp}$ during one gate is determined by the
ratio between operation time and lifetime of the states. The gate
time of the proposed gate is similar to the $\sigma^z$ geometric
phase gate with hyperfine states \cite{Leibfried}, the MS gate
\cite{Molmer-Sorensen, Haljan, Lee}, and the Cirac-Zoller
gate\cite{Cirac-Zoller, Schmidt-Kaler, Riebe} for the same laser
intensity, because in all these cases the coupling strength is
that of the first sideband interaction. The maximum intensity is,
however, limited by off-resonant excitations to $\Omega <
|\Delta_{1(2)}|\approx\nu/2$. The gate time $2 \pi/\delta=2
\pi/\frac{4 \eta \Omega^2}{\nu}$, therefore, is limited by $2
\pi/\eta\nu$. To guarantee that  $P_{\rm sp} < 10^{-4}$ , for
Ca${^+}$ ions the operation time has to be faster than 100 $\mu
s$. This means that it is necessary to increase $\Omega$ close to
the detunings $|\Delta_{1(2)}|$  and also to use a sufficiently
large Lamb-Dicke factor $\eta$ (see the dashed lines in
\ref{fig:etaomegadependency}).

When $\Omega$ is comparable  to the detunings, the direct coupling
term from off-resonant excitations, $\hat{H}_{d}=\sum_{j=1,2}
({\hat{\sigma}}_{1}^{+}+{\hat{\sigma}}_{2}^{+}) \frac{\hbar
\Omega_{j}}{2} e^{-i (\Delta_j t +\phi_j)}+ h.c. $, neglected in Eq.
(\ref{effectiveHamforone}) and (\ref{effectiveHamfortwo}), needs to
be considered. The term induces mainly population exchanges between
electronic states. 
The population error from off-resonant excitations can be
described by $1- (\frac{\Omega}{\nu/2})^2 {\sin}^2(\nu t/4) $
around $t=2\pi/\delta$ \cite{Molmer-Sorensen}. The infidelity can
be reduced by either precise control of system parameters like
$\delta= \nu/2N$, with $N$ an integer, or by pulse shaping. Using
pulse shaping, one can start and end the gate operation with a
fairly small $\Omega$ by adiabatically changing laser intensities
\cite{Wineland, Riebe,Roos08}. As we can see in Fig
\ref{fig:etaomegadependency}(b), the infidelity from the direct
coupling can be on the order of 10$^{-4}$ up to $\nu$/4 of
$\Omega$ using rise and fall times of 10 $\mu$s for the pulses.

In the simulation, we observed for a large $\Omega$ a reduction of
the Raman coupling strength $\Omega_{\rm eff}$ proportional to
$(\Omega/\nu)^2$. We believe that this reduction is due to an
admixture of the other electronic state due to for instance
off-resonant excitations. The Raman coupling $\Omega_{\rm eff,S}$ of
the $|S\rangle$ state has an opposite sign as compared to the
coupling $\Omega_{\rm eff,D}$ of the $|D\rangle$ state, $\Omega_{\rm
eff,S}=-\Omega_{\rm eff,D}$. Thus the contributions of other
electronic states
 due to off-resonant excitations reduce
the strength of the coupling $\Omega_{\rm eff}$ proportional to $(\Omega/\nu)^2$.

Furthermore, the amount of the reduction at $\Omega_{\rm eff,S}$
is slightly dependent on $\eta$, which might come from the
Debye-Waller factor \cite{Molmer-Sorensen}. We note that the
infidelities of Fig. \ref{fig:etaomegadependency}(a) and (b) are
obtained after correcting the reduction of $\Omega_{\rm eff}$. In
experiments, however, the change of $\Omega_{\rm eff}$ due to
direct coupling is not at all a problem, since the intensities of
the bichromatic lasers are to be determined anyways
experimentally.

We also carefully studied other experimental imperfections such as
 intensity fluctuations of both laser beams, positioning errors
of the laser beams on the ions, fluctuations of the laser and the
trap frequency, and the occupations of the bus mode and spectators'
modes. The proposed gate is quite robust to those imperfections
similar to the geometric hyperfine gate of Ref. \cite{Leibfried}.
According to our simulations, for intensity fluctuations of about
10$^{-2}$, a few tenths of kHz of laser frequency fluctuations, and
a few Hz trap frequency fluctuations, less than 0.5 motional quanta
of the all motional modes allow for an infidelity on a level of
10$^{-4}$.

In conclusion, we propose a $\sigma^z$ geometric phase gate for
optical transition qubits. The gate has a small spontaneous
emission during the operation, and can be applied to magnetic
field insensitive states. We analyze and simulate the gate in
detail and show that the gate allows to achieve a high fidelity
implementation. The proposed $\sigma^z$ gates are interesting not
only due to the high fidelity, but also to their applicability to
decoherence-free subspace constituted by the logical qubits
${|DS\rangle, |SD\rangle}$ \cite{Kielpinski, Aolita1}. If we apply
the laser beams described in the paper to two physical qubits that
belong to the different logical qubits, the scheme works as the
entangling gate for the two logical qubits.

We acknowledge support by the Austrian Science Fund (FWF), by the
European Commission (SCALA, CONQUEST networks), and by the
Institut f\"{u}ur Quanteninformation GmbH. KK acknowledges funding
by the Lise-Meitner program of the FWF. LA acknowledges the
support from CAPES, FAPERJ, and the Brazilian Millennium Institute
for Quantum Information.

\end{document}